\documentclass[conference,draftclsnofoot,onecolumn]{IEEEtran}
\IEEEoverridecommandlockouts

\usepackage{mathtools}
\mathtoolsset{showonlyrefs=true}

\usepackage{soul}
\usepackage[normalem]{ulem}

\usepackage[latin1]{inputenc}
\usepackage[english]{babel}
\usepackage{amsfonts,amssymb,amsmath,comment}
\usepackage[dvipsnames]{xcolor}
\usepackage{tikz}
\usetikzlibrary{decorations, decorations.text,backgrounds}
\usepackage{graphicx}
\usepackage{subfig}

\usepackage{cite}
\usepackage{placeins}

\renewcommand{\natural}{{\mathbb{N}}}

\newcommand{\real}{{\mathbb{R}}}

\newcommand{\norm}[1]{\|#1\|}

\newcommand{\until}[1]{\{1,\ldots,#1\}}

\newcommand{\EE}{\mathcal{E}} 
\newcommand{\GG}{\mathcal{G}}

\newcommand{\KK}{\mathcal{K}}
 
\newcommand{\NN}{\mathcal{N}} 
\renewcommand{\SS}{\mathcal{S}}

\newcommand{\argmin}{\mathop{\rm argmin}}

\newcommand{\nbrs}{\mathcal{N}}

\usepackage{algorithm}
\usepackage[noend]{algpseudocode}

\makeatletter
\newcommand{\StatexIndent}[1][3]{%
  \setlength\@tempdima{\algorithmicindent}%
  \Statex\hskip\dimexpr#1\@tempdima\relax}
\makeatother

\algnewcommand{\algorithmicgoto}{\textbf{go to }}%
\algnewcommand{\Goto}[1]{\algorithmicgoto Line~\ref{#1}}%
\algnewcommand{\Label}{\State\unskip}

\newtheorem{theorem}{Theorem}[section]

\newtheorem{remark}[theorem]{Remark}

\newtheorem{assumption}[theorem]{Assumption}

\renewcommand{\lim}{\operatornamewithlimits{lim\vphantom{p}}}

\newcommand\oprocendsymbol{\hbox{$\square$}}
\newcommand\oprocend{\relax\ifmmode\else\unskip\hfill\fi\oprocendsymbol}

\graphicspath{{figs/}}

\newcommand{\bx}{\mathbf{x}}

\newcommand{\hbu}{\widehat{\mathbf{u}}}
\newcommand{\hbg}{\widehat{\mathbf{g}}}

\newcommand{\avg}{\bar{\mathbf{x}}}
\newcommand{\bv}{\mathbf{v}}
\newcommand{\bu}{\mathbf{u}}
\newcommand{\bn}{\mathbf{n}}

\newcommand{\by}{\mathbf{y}}
\newcommand{\bA}{\mathbf{A}}
\newcommand{\bD}{\mathbf{D}}
\newcommand{\bb}{\mathbf{b}}

\newcommand{\1}{\mathbf{1}}
\newcommand{\bpi}{\boldsymbol{\pi}}

\newcommand{\bphi}{\boldsymbol{\phi}}

\newcommand{\tf}{\tilde{f}}

\definecolor{blue@O4S}{RGB}{0, 41, 69}
\definecolor{emph@O4S}{RGB}{0, 93, 137}
\definecolor{red@O4S}{RGB}{127,0,0}
\definecolor{gray@O4S}{RGB}{112, 112, 112}

\def \algfullname/{{\sc Block-SONATA}}
\def \algacronym/{{\sc Block-SONATA}}

\begin{document}
\title{Distributed Big-Data Optimization  via Block Communications}

\author{Ivano Notarnicola$^{\ast}$, Ying Sun$^{\ast}$, Gesualdo Scutari, Giuseppe Notarstefano
\thanks{$^\ast$These authors equally contributed and are in alphabetic order.}
\thanks{The work of Notarnicola and Notarstefano has received funding from the
  European Research Council (ERC) under the European Union's
  Horizon 2020 research and innovation programme (grant agreement No 638992 -
  OPT4SMART).
  The work of Sun and Scutari was supported by the USA NSF Grants CIF 1632599, CIF 1719205, and CAREER Award 1555850; and in part 
  by the ONR Grant N00014-16-1-2244.
  }
\thanks{Ivano Notarnicola and Giuseppe Notarstefano are with the
  Department of Engineering, Universit\`a del Salento, Lecce, Italy, 
  \texttt{name.lastname@unisalento.it.}  } 
\thanks{Ying Sun and Gesualdo Scutari are with the School of Industrial Engineering, 
Purdue University, West-Lafayette, IN, USA, \texttt{\{sun578,gscutari\}@purdue.edu.}}}

\maketitle

\begin{abstract}
  We study  distributed multi-agent large-scale optimization problems, wherein the cost function is composed of
a smooth possibly nonconvex sum-utility  plus  a DC (Difference-of-Convex) regularizer. 
We consider the scenario where the dimension of the optimization variables is so large that optimizing and/or transmitting the entire set of variables could cause unaffordable computation and communication overhead. 
To address this issue, we propose   the first  distributed algorithm  whereby  agents optimize and communicate   only a portion of their local variables.
The scheme hinges on  successive convex approximation (SCA) to handle the nonconvexity of the objective function, coupled with  a novel block-signal tracking scheme, aiming at 
locally estimating the average of the agents' gradients.
 Asymptotic convergence to
stationary solutions of the nonconvex problem is established. Numerical results on  a sparse regression problem show   
  the effectiveness of the proposed algorithm and the impact of the block size on its practical convergence speed and communication cost.
\end{abstract}
\IEEEpeerreviewmaketitle

\section{Introduction}
We consider a multi-agent system composed of $N$ agents that cooperatively aim at solving the following (possibly nonconvex)  optimization problem:
\begin{equation}\label{prob:P}\tag{P}
\begin{aligned}
& \underset{\bx }{\text{minimize}} && U(\bx)\triangleq \sum_{i=1}^N f_i(\bx) + \sum_{\ell=1}^B \underbrace{r_\ell^+(\bx_\ell) - r_\ell^-(\bx_\ell) }_{r_\ell({\bx_\ell})}\\
& \text{subject to} && \bx = [\bx_1^\top,\ldots,\bx_B^\top]^\top\\ 
&                      && \bx_\ell \in \KK_\ell, \quad\forall \ell \in \until{B},
\end{aligned}
\end{equation}
where $\bx\in\real^{dB}$ is the vector of the optimization variables,  partitioned in $B$ blocks, whose   $\ell$-th block is  denoted by $\bx_\ell \in \real^d$;
 $f_i: \real^{dB} \to \real$ is a smooth possibly nonconvex   cost function of agent $i$;  
$r_\ell:\real^d \to \real$, $\ell \in \until{B}$, is a difference of convex (DC) function commonly known by all the agents; %
 and $\KK_\ell$, $\ell \in \until{B}$, is a  closed  convex set.
Function $r_\ell$ usually plays the role of  a regularizer, used to  promote some favorable structure on the solution $\bx$, such as sparsity. 
The DC structure of  $r_\ell$ is motivated by the need of capturing in a unified formulation both convex and  nonconvex regularizers, the latter being   shown  to achieve superior performance than their convex counterparts \cite{pang2016PartI}.
Problem \eqref{prob:P} is of broad interest and models a wide range of applications including network resource allocation, target localization, as well as statistical learning problems.

Our goal is to design a distributed algorithm  solving large-scale instances of \eqref{prob:P}.
These problems, also referred to as big-data problems, pose the following two challenges:
(i) optimizing the objective function, or even just computing the gradient  with respect to all the variables, can be too costly; 
(ii)  broadcasting over the network at each iteration all agents' local variables   would incur in an unaffordable communication overhead.
We are not aware of any work in the literature that can address both challenges (i) and (ii) for problem~\eqref{prob:P}. In fact, as discussed next,  the existing distributed algorithms  either call for the  optimization  and transmission of  the entire vector  $\bx$ per iteration (or auxiliary variables of the same size of $\bx$)  
or impose restrictive structures on the objective function to work. %

There is a vast literature of  distributed algorithms for  both  convex \cite{nedic2015distributed,nedic2010constrained,tsitsiklis1986distributed,chen2012fast,duchi2012dual,boyd2011distributed,mota2013d,shi2014linear,chang2015multi} and nonconvex problems \cite{bianchi2011convergence,tatarenko2015non,hong2016decomposing,bianchi2013performance,dilorenzo2016next,sun2017distributed}. 
Although  substantially different,  these methods are all based on two main steps, namely: a local optimization and then a communication step of the \emph{entire} vector $\bx$   (or some related variables  of the same size, e.g., multipliers). %
They thus fail to address challenges (i) and (ii).
 On the other hand, (block) coordinate descent methods \cite{wright2015coordinate,tsitsiklis1986distributed,nesterov2012efficiency,mokhtari2016doubly} and parallel algorithms \cite{richtarik2012parallel,richtarik2014iteration,necoara2016parallel,facchinei2015parallel}     
 have been shown to be quite effective in handling large-scale problems by optimizing one block of the variables per time. 
These algorithms, however, are not readily implementable in the aforementioned distributed setting, because they   assume that either all  agents know the whole sum-utility  or that, at each iteration, each agent has access to the current value of the other agents' variables. 
While these assumptions are naturally satisfied in a share-memory system (e.g., data centric architecture) or complete (graph) networks, if enforced for  problem~\eqref{prob:P}, they would call for an  heavy message passing among the agents. %
We are aware of  only a few distributed schemes operating on block  variables, namely:  \cite{notarnicola2016randomized,carli2013distributed,recht2011hogwild}. They however   
 require a certain degree of graph separability on the sum-utility function, meaning that  each agent's function $f_i$ can depend only on the  variables of that agent and its neighbors, which makes them  not applicable to problem~\eqref{prob:P}.

In this work, we propose   \algacronym/, the first distributed algorithm for the general class of  problem~\eqref{prob:P} that is able to address both challenges (i) and (ii): %
each agent   iteratively optimizes and transmits only one block of its local variables.
More specifically,  \algacronym/ consists of two steps, namely: 1) a local optimization step wherein agents locally solve a   covexification of \eqref{prob:P}, with respect to a chosen block of their local variables; and 2) 
 a \emph{blockwise} consensus step, aiming at forcing an agreement among the agents' local copies. 
Moreover,   a novel blockwise signal tracking scheme is also  employed   to dynamically estimate  the gradient of the sum-utility function, using only local information via block-communications.  
Agents select the blocks to optimize/transmit in an uncoordinated fashion.  %
Asymptotic convergence  is established under   mild   assumptions. Compared to our recent proposal \cite{notarnicola2017distributed},  \algacronym/ is computationally more efficient, since it does not require at each iteration the computation of the entire gradient of the functions  $f_i$.

 \section{Problem Setup}
We study problem~\eqref{prob:P} under the following assumptions.

\begin{assumption}[On Problem~\eqref{prob:P}] 
\begin{enumerate} 
  \item $\KK_\ell \neq \emptyset$  is closed and convex;
  \item $f_i:\mathbb{R}^{dB}\rightarrow \mathbb{R}$  is $C^1$ on (an open set containing)  $\KK$;
  \item $\nabla f_i$ is $L_i$-Lipschitz  continuous 
    and bounded on $\KK$;
  \item $r_\ell^+:\mathbb{R}^d\rightarrow \mathbb{R}$ is convex (possibly nonsmooth)  on $\KK$,
  with bounded subgradients on $\KK$; and $r_\ell^- :\real^d \to \real$ is convex on $\KK$, with   Lipschitz continuous bounded gradient  on $\KK$;
  \item $U$ is coercive on $\KK$, i.e., $\lim_{\bx\in\KK,\|\bx\|\to\infty} U(\bx) = \infty$. 
\end{enumerate}
\label{ass:cost_functions}
\end{assumption}
Assumption \ref{ass:cost_functions} is standard and can be easily satisfied in
practice; see, e.g.,  \cite{facchinei2015parallel}. Here we only  remark that both the local cost $f_i$
and the common regularizer $\sum_{\ell=1}^B r_\ell$ need not be convex. \smallskip

\noindent \emph{On the communication
  network:}
The communication among the agents is modeled by a fixed, directed graph
$\GG = (\until{N},\EE)$, where $\EE\subseteq \until{N} \times \until{N}$ is the
set of edges. There is an edge $(i,j)\in \EE$ if agent $i$ can send a message to agent
$j$.
We denote by $\nbrs_i$ the set of \emph{in-neighbors} of node $i$ in $\GG$, including itself, i.e.,
$\nbrs_i \triangleq \left\{j \in \until{N} \mid (j,i) \in \EE \right\}\cup \{i\}$. %
To let information propagate over the network, we make the following
assumption. \smallskip
\begin{assumption}[Network connectivity]
The graph $\GG$ is strongly connected.
\label{ass:strong_conn}
\end{assumption} 
\smallskip

The above setting and problem  are quite general and model  many 
applications of practical interest. An example in the context of sparse  signal estimation is briefly outlined next.

\noindent \texttt{Sparse regression:} Consider the problem of estimating a sparse signal $\bx_0$ from linear measurements $\{\bb_i\}_{i=1}^N$, where 
$\bb_i = \bD_i\bx_0 + \bn_i$ with $\bn_i$ being the measurement noise at agent $i$'s side. The problem can be formulated as
\begin{equation}
  \begin{aligned}
  & \underset{\bx \in \KK  }{\text{minimize}} &&  \sum_{i=1}^N \|\bb_i - \bD_i \bx\|^2+ R(\bx),
  \end{aligned}
\label{prob:sparse_regression}
\end{equation}
where $R:\real^{dB}\to\real$ is a  sparsity-promoting regularizer having the structure
$
 R(\bx) \triangleq \lambda\cdot \sum_{k=1}^{dB} r(x_k)
$ [cf.~(P)], with $\lambda>0$. 
The DC structure of $R$ is motivated by the fact that both  convex regularizers (e.g., $\ell_1$, $\ell_2$, and  elastic net) and the widely used nonconvex regularizers (e.g., SCAD, Log, Exp, $\ell_p$ norm for $0<p<1$) can be written as  \cite{pang2016PartI}
\begin{equation}
r(x) \triangleq \underbrace{\eta(\theta)\cdot  |x|}_{r^+(x)} - \underbrace{\left(\eta(\theta)\cdot  |x| - r(x)\right)}_{r^-(x)}, \label{eq:DC-reg}
\end{equation}
 where $r^-:\real \to \real$ is a convex function with Lipschitz continuous derivative. Problem \eqref{prob:sparse_regression} is clearly an instance of Problem \eqref{prob:P}.

\section{Algorithmic Design}

Before describing the proposed distributed algorithm, we introduce a block-wise dynamic average
consensus scheme, whereby the agents aim at cooperatively tracking the average
 of a time-varying signal via block-wise communications.

\subsection{Average signal tracking via block communications}

We consider the problem of tracking the average of a signal  over a graph $\GG$ satisfying Assumption~\ref{ass:strong_conn}.
Each agent $i$ can evaluate locally a time-varying signal $\{\bu_i^t\}_{t\in\natural}$, and all agents aim at tracking the average signal $\bar{\bu}^t\triangleq \frac{1}{N}\sum_{i=1}^N \bu_i^t$ by exchanging information over the network.
We assume that the cost of acquiring  $\bu_i^t$ is non-negligible, e.g., 
$\bu_i^t$ can be the gradient of a function with respect to a large number of variables. Distributed tracking has been studied in  \cite{sun2017distributed}. However, such a scheme requires at each iteration  the acquisition  of the \emph{entire} signal $\bu_i^t$ as well as the communication of a vector having the same size of $\bu_i^t$, which is too costly.  To cope with the curse of dimensionality,  we develop next a signal tracking scheme that %
operates at the level of the blocks of signals $\bu_i^t$ while enabling block-communications.

Each agent $i$ maintains a local variable $\bx_{(i)}^t$, whose $\ell$-th block is denoted by  $\bx_{(i,\ell)}^t$, with  $\ell \in \until{B}$. 
At   iteration $t$, each agent $i$    picks a block-index, say  $\ell_i^t$, and broadcasts the block $\bx_{(i,\ell_i^t)}^t$ to its neighbors.  Based 
on the information (blocks) received from its neighbors and the acquired block of the local signal $\bu_{i}^t$, agent $i$  updates block-wise its entire vector $\bx_{(i)}^t$ (according to the mechanism that we will introduce shortly).
Since there is no coordination among the agents, they will likely transmit blocks associated with different indices. This implies that blocks with different index will ``travel'' on different communication graphs, which in general do not coincide with $\GG$: 
agent $j (\neq i)$ is an in-neighbor of $i$ if 
$j\in \nbrs_i$ and agent $j$ sends block $\ell$ to $i$ at iteration $t$. This naturally suggests the adoption of 
  block-dependent  communication graphs, one per block $\ell$. Specifically,   
$\GG_\ell^t\triangleq (\until{N},\EE_\ell^t)$, which is a  \emph{time-varying} 
subgraph of $\GG$ associated to block $\ell$ at iteration $t$, whose edge set   is defined as 
$
  \EE_\ell^t \triangleq \{(j,i)\in\EE \mid j\in \nbrs_{i,\ell}^t, i\in\until{N}\},
$
where $\nbrs_{i,\ell}^t$ is the   in-neighborhood  of agent $i$ associated with the block-index  $\ell$, 
$
  \nbrs_{i,\ell}^t \triangleq \{ j \in \nbrs_i \mid \ell_j^t = \ell \}  \cup \{i \} \subseteq \nbrs_i.
$

Using block-dependent graphs one can solve the tracking problem block-wise. Therefore, in the following, 
we focus only on block $\ell$, without loss of generality. 
The task reduces to developing a tracking algorithm over the time-varying directed graph $\{\GG_\ell^t\}_{t\in\natural}$. 
Building on \cite{sun2017distributed}, we propose the following adapt-then-combine scheme:
\begin{align}
\begin{split}
  \bv_{(i,\ell)}^{t} & = \bx_{(i,\ell)}^{t} + \frac{1}{\phi_{(i,\ell)}^t}(\bu_{i,\ell}^{t+1} - \bu_{i,\ell}^t)
  \\[0.2cm]
  \phi_{(i,\ell)}^{t+1} & = \sum_{j \in \nbrs_{i,\ell}^t} a_{ij\ell}^t\, \phi_{(j,\ell)}^t, \quad \phi_{(i,\ell)}^0 = 1,\hspace{.5cm}  \ell \in \until{B},
  \\
  \bx_{(i,\ell)}^{t+1} & =\frac{1}{\phi_{(i,\ell)}^{t+1}}\,\sum_{j \in \nbrs_{i,\ell}^t} a_{ij\ell}^t \,\phi_{(j,\ell)}^t\, \bv_{(j,\ell)}^t,
\end{split}
\label{eq:block_tracking}
\end{align}
where $\{a_{ij\ell}^t\}_{ij}$ is a set of weights that need to be properly chosen. Collecting these weights in a matrix $\bA_\ell^t\triangleq [a_{ij\ell}^t]_{ij}$,  we make the following standard assumptions on $\bA_\ell^t$.
\smallskip 
\begin{assumption}[On the Weighting Matrix $\bA_\ell^t$]
  For all $\ell \in \until{B}$ and $t>0$, matrix $\bA_\ell^t$ satisfies the following conditions:
  \begin{enumerate}
    \item $a_{ii\ell}^t \geq \vartheta >0$, for all $i\in\until{N}$;
    \item $a_{ij\ell}^t \geq \vartheta >0$, for all $(j,i) \in \EE_\ell^t$;
    \item $\bA_\ell^t$ is column stochastic, i.e., $\1^\top \bA_\ell^t = 
    \1^\top$.
 \end{enumerate}
  \label{ass:col_stoch}
\end{assumption}
\smallskip 

Roughly speaking, the block-tracking scheme in (\ref{eq:block_tracking}), can be interpreted as follows:   each agent first updates its local estimate towards the current signal 
$\bu_{i,\ell}^{t+1}$, and then averages it with the local updates  of 
its neighbors. 
The  scalar variable $\phi_{(i,\ell)}$ is introduced to obtain a convex combination of the received   $\bv_{(i,\ell)}$'s through the equivalent weights $(a_{ij\ell}^t \,\phi_{(j,\ell)}^t)/\phi_{(i,\ell)}^{t+1}$ (recall that, by Assumption~\ref{ass:col_stoch}, $\bA_\ell^t$ is column stochastic, but in general is not row stochastic).

While the tracking scheme \eqref{eq:block_tracking} unlocks block-communications, it still  requires, at each iteration, the acquisition  of the entire signal $\bu_i^t$.  To cope with this issue, we propose to replace $\bu_i^t$ with  a surrogate local variable, denoted by 
$\hbu_i^t$,  initialized as
$\hbu_i^0 = \bu_i^0$.  At iteration $t$, agent $i$ acquires only a block of signal
$\bu_i^{t}$, say block $\ell_i^t$  for notation simplicity, and updates $\hbu_i^t$  as
\begin{equation}
  \hbu_{i,\ell}^{t} = 
  \begin{cases}\label{u_hat_update}
    \bu_{i,\ell}^{t}, & \text{if } \ell = \ell_i^t,
    \\
    \hbu_{i,\ell}^{t-1}, & \text{if } \ell \neq \ell_i^t,
  \end{cases}
\end{equation}
where $\bu_{i,\ell}^t$ [resp. $\hbu_{i,\ell}^t$] denotes the $\ell$-th block of 
$\bu_i^t$ [resp. $\hbu_{i}^t$]. 
That is, vector $\hbu_i^t$ collects agent $i$'s most 
recent information on   $\bu_i^t$.

To summarize, the proposed block-tracking scheme reads as \eqref{eq:block_tracking}, where  $\bu_{i,\ell}^{t}$ [resp. $\bu_{i,\ell}^{t+1}$] is replaced by $\hbu_{i,\ell}^{t}$ [resp. $\hbu_{i,\ell}^{t+1}$], defined in \eqref{u_hat_update}.
To ensure  convergence--i.e.,  $\lim_{t\to\infty}\|\bx_{(i,\ell)}^t - \hbu_\ell^t\| = 0$, for all $\ell$--we need the following  assumptions on the connectivity of $\{\GG_\ell^t\}_{t\in\natural}$, which is  widely used in the literature of push-sum-like algorithms.\smallskip 
\begin{assumption}
For all $\ell\in \until{B}$, there exists a finite integer $T>0$ such that the graph   sequence $\{\GG_\ell^t\}_{t\in\natural}$ is $T$-strongly connected, i.e.,  the union graph 
$(\until{N},\cup_{s=t}^{t+T-1}\,\EE_\ell^s)$ is strongly connected, for all $t>0$. \label{ass:T_connectivity}
\end{assumption}
 Since each  digraph $\GG_\ell^t$ is induced by the adopted  block-selection rule, its connectivity clearly depends on it.
 A key  question, addressed next,  is then: how to design, in a distributed and uncoordinated way,  agents' block-selection rules and $\bA_\ell^t$  that fulfill Assumption    \ref{ass:col_stoch} and \ref{ass:T_connectivity}? 
 
By the definition of $\GG_\ell^t$, all the  edges in the underlying graph $\GG$ leaving node $i$ will be also edges of $\GG_\ell^t$ if agent $i$ sends block $\ell$ at time $t$.
Since   $\GG$ is strongly connected (cf.\,Assumption\,\ref{ass:strong_conn}), $\GG_\ell^t$ is $T$-strongly connected if, starting from any time $t>0$, all   agents send block $\ell$ within $T$ iterations,
which translates in the following essentially cyclic block-selection rule.\smallskip 
\begin{assumption}[Block-selection Rule]
	For each 	agent $i\in\until{N}$, there exists a (finite) constant $T_i>0$ such that
	$
	  \cup_{s=0}^{T_i-1} \{\ell_i^{t+s} \} = \until{B}, \text{ for all } t \ge 0. 
  $
\label{ass:block_selection}
\end{assumption}
\smallskip 
Note that the above rule does not impose any coordination among the agents: at each iteration,  
  different agents may update different blocks. 
It is not difficult to  show that, under Assumptions~\ref{ass:strong_conn} and \ref{ass:block_selection}, 
there exits a  $0<T\leq \max\limits_{i\in\until{N}} \!T_i$, such that 
$\cup_{s = 0}^{ T-1} \GG_\ell^{t+s}$, $\ell\in \{1,\ldots, B\}$, is strongly connected, for all $t\geq 0$.

We show next  how agents can \emph{locally} build a matrix $\bA_\ell^t$ satisfying 
Assumption~\ref{ass:col_stoch}. %
Observe that at iteration $t$, if agent $j$ selects block $\ell$, it sends $\bv_{(j,\ell)}^t$ 
to any agent $i$ that is its out-neighbor; or send it to no one, otherwise.
In addition, $a_{jj\ell}^t$ must be nonzero by Assumption~\ref{ass:col_stoch}.
Consequently, the $j$-th column of $\bA_\ell^t$, denoted by $\bA_\ell^t(:,j)$, can only have the following two possible sparsity patterns:
(i) all $a_{ij\ell}^t$, with $i\in \{\until{B}:(j,i)\in \EE\}$, is nonzero if $\ell_j^t = \ell$;
(ii) only $a_{jj\ell}^t$ is nonzero if $\ell_j^t \neq \ell$.
To meet the requirement that $\bA_\ell^t$ is column stochastic, 
agent $j$ thus either  select a stochastic vector  $\bA_\ell^t(:,j)$ matching 
the  sparsity pattern described in case (i), if $\ell_j^t = \ell$; or set $\bA_\ell^t(:,j)$ 
to be the  $j$-th vector of the canonical basis, if $\ell_j^t \neq \ell$. It is not difficult to check that such   weights   can be  constructed locally by each agent, with no coordination with the others. 

We conclude this section,  noting that the proposed block-trackig scheme can be used also to solve the  average consensus problem  wherein   agents aim to estimate the average of 
their initial estimates, i.e., $({1}/{N})\sum_{i=1}^N\bx_{(i)}^0$. Specifically, by reinterpreting the 
consensus problem as tracking of the average of 
the constant signal $\bx^0 \!\!\triangleq \!\![\bx_{(1)}^{0\top},\ldots,\bx_{(N)}^{0\top}]^\top\!\!,\!$ it is enough 
to  set   $\bu_i^t \!\equiv \!\bx^0$, $t\!\geq\! 0$, and absorb the $\bv$-variable, which leads to  the 
following \emph{block-consensus} algorithm:
\begin{equation}\label{eq:block_consensus}
 \begin{array}{ll}
  \phi_{(i,\ell)}^{t+1} & = \displaystyle{\sum_{j \in \nbrs_{i,\ell}^t} }a_{ij\ell}^t \phi_{(j,\ell)}^t, \medskip 
  \\
 \bx_{(i,\ell)}^{t+1} & = \dfrac{1}{\phi_{(i,\ell)}^{t+1}}\displaystyle{\sum_{j \in \nbrs_{i,\ell}^t}} a_{ij\ell}^t \phi_{(j,\ell)}^t \bx_{(j,\ell)}^t,
 \end{array}\quad \forall \ell \in \until{B}.
 \end{equation}

\subsection{\algfullname/: A constructive approach}
We are now in the position to introduce our algorithmic framework.
Observe that what couples the functions $f_i$ in Problem~\eqref{prob:P} is the common vector variable 
$\bx$. To decouple the problem 
a natural step is  then introducing for each agent $i$ a local copy  $\bx_{(i)}$ of  $\bx$. 
Yet, agent $i$ faces the following challenges: (i) the dimension of $\bx_{(i)}$ is large; (ii) $f_i$ and $-r_\ell^-$ 
are nonconvex; and (iii) $\sum_{j\neq i} f_j$ is unknown.
To cope with these issues, we introduce \algacronym/ (cf.~Algorithm 1),  an iterative scheme leveraging   SCA   
 techniques, coupled with  a parallel blockwise consensus/tracking step based on \eqref{eq:block_tracking} 
 and \eqref{eq:block_consensus}, as detailed next.
\begin{algorithm}[!ht]
\renewcommand{\thealgorithm}{}
\floatname{algorithm}{Algorithm 1:}
  \begin{algorithmic}
    \StatexIndent[0] Set $t=0$, $\bphi_{(i)}^0 = \1$, $\hbg_i^0 = \by_{(i)}^0 = \nabla f_i(\bx_{(i)}^0)$, $\ell_i^0 \in \until{B}$. 
    \medskip
      
    \StatexIndent[0] \textbf{Local Optimization}: 
      \begin{align}      
            \widetilde{\bpi }_{(i,\ell_{i}^{t})}^{t} = & N \cdot \by_{(i,\ell_i^t)}^{t} - \nabla_{\ell_i^t} f_i \big( \bx_{(i)}^{t} \big),
    \label{eq:alg_pi_update}\\ 
\widetilde{\bx}_{(i,\ell_i^t)}^t \triangleq & \argmin_{\bx_{\ell_i^t} \in \KK_{\ell_i^t}} \ r^+_{\ell_i^t}(\bx_{\ell_i^t}) + \widetilde{f}_{i,\ell_i^t}(\bx_{\ell_i^t};\bx_{(i)}^t)
  + (\widetilde{\bpi}_{(i,\ell_i^t)}^t - \nabla r^-_{\ell_i^t}(\bx_{(i,\ell_i^t)}^t))^\top (\bx_{\ell_i^t} - \bx_{(i,\ell_i^t)}^t),\label{eq:alg_x_min} \\
         \bv_{(i,\ell_i^t)}^{t} = & \ \bx_{(i,\ell_i^t)}^t + \gamma^t ( \widetilde{\bx}_{(i,\ell_i^t)}^t - \bx_{(i,\ell_i^t)}^t );\label{eq:alg_v}
         \end{align}
      \StatexIndent[0.25] Broadcast $\bv_{(i,\ell_i^t)}^t$,
      $\phi_{(j,\ell)}^t$, $\by_{(j,\ell_i^t)}^t$ to the
      out-neighbors; %

    \medskip
    
    \StatexIndent[0] \textbf{Averaging and Gradient Tracking}: 
    \StatexIndent[0] For \!$\ell \!\in \!\until{B}$:\!
   receive $\phi_{(j,\ell)}^t$,$\bv_{(j,\ell)}^t$ %
     from $j \!\in\! \nbrs_{i,\ell}^t$, and set
    \begin{align}
      \phi_{(i,\ell)}^{t+1} & = \sum_{j \in \nbrs_{i,\ell}^t} a_{ij\ell}^t \,\phi_{(j,\ell)}^t,
    \label{eq:alg_phi_update}
    \\
      \bx_{(i,\ell)}^{t+1} & = 
      \dfrac{1}{\phi_{(i,\ell)}^{t+1}} \sum_{j\in \nbrs_{i,\ell}^t } a_{ij\ell}^t\, \phi_{(j,\ell)}^t \bv_{(j,\ell)}^t;
    \label{eq:alg_x_update}      
    \end{align}
      \StatexIndent[0]    Select $\ell_i^{t+1} \in \until{B}$ and update
      \begin{align}
          \hbg_{i,\ell}^{t+1} & = 
    \begin{cases}
    \nabla_{\ell_i^{t+1}}  f_i(\bx_{(i)}^{t+1}),
    & \text{ if } \ell = \ell_i^{t+1},
    \\
    \hbg_{i,\ell}^{t}, & \text{ otherwise};
    \end{cases}    \label{eq:alg_g_update}
    \end{align}
        \StatexIndent[0] For  $\ell \!\in \!\until{B}$: 
    receive $\phi_{(j,\ell)}^t \,\by_{(j,\ell)}^t + \hbg_{j,\ell}^{t+1}
       -
        \hbg_{j,\ell}^{t}$ from $j \!\in\! \nbrs_{i,\ell}^t$, and set
    \begin{align}   
        \begin{split}
      \by_{(i,\ell)}^{t+1} & = 
      \dfrac{1}{\phi_{(i,\ell)}^{t+1}} 
      \sum_{j \in \nbrs_{i,\ell}^t}
      \!\!
       a_{ij\ell}^t\, \Big(\phi_{(j,\ell)}^t \by_{(j,\ell)}^t 
      + \hbg_{j,\ell}^{t+1}
       -
        \hbg_{j,\ell}^{t}\Big).    \label{eq:alg_y_update}
            \end{split}
      \end{align}

  \end{algorithmic}
  \caption{\algfullname/}
  \label{alg:algorithm}
\end{algorithm}

\noindent\textit{Local optimization:}
At iteration $t$, agent $i$  selects and optimizes   a block of $\bx_{(i)}^t$, say $\ell_i^t$ [this addresses challenge (i)].
To deal with the nonconvexity of $f_i$ and $-r_{\ell_i^t}^-$ [challenge (ii)], 
we approximate $f_i$ with a strongly convex surrogate $\tf_{i,\ell_i^t}$ and   $-r_{\ell_i^t}^-$ by its linearization 
at $\bx_{(i,\ell_i^t)}^t$.
The unknown term 
$\sum_{j\neq i}f_j$ is replaced by a linear function whose coefficient $\widetilde{\bpi}_{(i,\ell_i^t)}^t$ aims to track $\sum_{j\neq i}\nabla f_{j,\ell_i^t}(\bx_{(i)}^t)$ [challenge  (iii)].
The resulting problem~\eqref{eq:alg_x_min}  is thus a strongly convex approximation of problem~\eqref{prob:P} and admits a unique solution $\widetilde{\bx}_{(i,\ell_i^t)}^t$.
Agent $i$ then updates  its local copy $\bx_{(i,\ell_i^t)}^t$ along direction $\widetilde{\bx}_{(i,\ell_i^t)}^t - \bx_{(i,\ell_i^t)}^t$,  
with step-size $\gamma^t$, see~\eqref{eq:alg_v}. Note that agent $i$ does not optimize blocks $\ell \neq \ell_i^t$, hence we let $\bv_{(i,\ell)}^t = \bx_{(i,\ell)}^t,\ \forall \ell \neq \ell_i^t$.  
Agent $i$ then broadcasts $\bv_{(i,\ell_i^t)}^t$ to its neighbors. \smallskip

\noindent\textit{Blockwise consensus/gradient tracking:} To force consensus on $\bx_{(i)}^t$, agent $i$ update in parallel all its blocks $\bx_{(i,\ell)}^t$, $\ell \in \until{B}$, based on the received variables $\bv_{(j,\ell_j^t)}^t$. Leveraging  the block consensus scheme~\eqref{eq:block_consensus}, the aforementioned updates read~\eqref{eq:alg_phi_update}--\eqref{eq:alg_x_update}. 

Finally, we need to introduce the update of $\widetilde{\bpi}_{(i,\ell)}^t$ so that 
$\lim_{t\to\infty}\|\widetilde{\bpi}_{(i,\ell)}^t - \sum_{j\neq i} \nabla_\ell f_{j}(\bx_{(i)}^t)\|=0$.
To this end, we rewrite $\sum_{j\neq i}\nabla_\ell f_{j}(\bx_{(i)}^t)$ as 
\begin{align*}
  \sum_{j\neq i}
  \nabla_\ell f_{j}(\bx_{(i)}^t) = 
  N\cdot \underbrace{\frac{1}{N}\sum_{j=1}^N \nabla_\ell f_{j}(\bx_{(i)}^t)}_{\overline{\nabla_\ell f}(\bx_{(i)}^t)} 
  \,-\, \nabla_\ell f_{i}(\bx_{(i)}^t).\\[-.7cm]
\end{align*}
Since $\nabla_\ell f_{i}(\bx_{(i)}^t)$ can be evaluated locally by agent $i$, the task boils down 
to estimate the average gradient $\overline{\nabla_\ell f}(\bx_{(i)}^t)$, $\forall \ell \in \until{B}$.
We can then readily invoke the blockwise tracking scheme~\eqref{eq:block_tracking} with $\ell_i^{t+1}$ selected according to the essentially cyclic rule (Assumption~\ref{ass:block_selection}) and  $\bu_{i}^t \triangleq \nabla f_i (\bx_{(i)}^t)$, leading to the updates \eqref{eq:alg_g_update}-\eqref{eq:alg_y_update}.

\begin{remark}
  Note that the block selected in the tracking step~\eqref{eq:alg_g_update} needs not to 
  be the same as the one used in the optimization step~\eqref{eq:alg_pi_update}. However, 
  in \algacronym/ we let them be equal so that to perform the two aforementioned steps only 
  one block-gradient computation is needed.\oprocend
\end{remark}

Having introduced the algorithm, the remaining question is how to choose the surrogate functions $\tilde{f}_i$ and the step-size $\gamma^t$.  
Convergence of \algacronym/ is guaranteed under the following assumptions.
\begin{assumption}[On the Surrogate Functions]
Given Problem~\eqref{prob:P} under Assumption~\ref{ass:cost_functions}, 
each  surrogate function 
$\tf_{i,\ell}: \KK_\ell\times \KK\rightarrow \real$ satisfies: %
\begin{enumerate}
\item 
  $\tf_{i,\ell} (\bullet;\bx)$ is uniformly strongly convex  on $\KK_\ell$;%
\item
  $\nabla \tf_{i,\ell} (\bx_{\ell};\bx) = \nabla_\ell f_i (\bx)$, for all $\bx \in \KK$;
\item
  $\nabla \tf_{i,\ell} (\bx_{\ell}; \bullet)$ is uniformly Lipschitz continuous on $\KK$;
\end{enumerate}
where $\nabla \tf_{i,\ell}$ denotes the partial gradient of $\tf_{i,\ell}$ 
with respect to its first argument.
\label{ass:surrogate}
\end{assumption}

\begin{assumption}[On the step-size]
The sequence $\{\gamma^t\}$, with each $0< {\gamma^t} \le 1$, satisfies:
(i) $\sum\limits_{t=0}^{\infty}\gamma^t = \infty$ and $
  \sum\limits_{t=0}^{\infty} (\gamma^t)^2 < \infty$; 
  (ii) $\gamma^t/\eta \le  \gamma^{t+1} \le \gamma^t$,
  for all $t\geq 0$ and  some $\eta\in (0,1)$.
\label{ass:step-size}
\end{assumption}

Assumption~\ref{ass:surrogate} states that  $\tilde{f}_i$ should be regarded as a (simple) strongly convex approximation of $f_i$
that  preserves its first order properties. Several valid choices for $\tilde{f}_i$ are available; see, e.g., \cite{facchinei2015parallel,dilorenzo2016next}.
Assumption~\ref{ass:step-size} is the standard diminishing step-size rule (i) with 
the extra requirement (ii), which  ensures all the blocks contribute ``equally'' to the optimization.
Condition (ii) can be met easily in practice \cite{cannelli2016asynchronous,cannelli2017asynchronous}; an example is given in Sec.~\ref{sec:simulations}.
The convergence of \algacronym/ is given in the following theorem, whose proof is omitted 
due to space limitation; see~{\cite{notarnicola2017technical}}.%

\begin{theorem}
  Let  $\{(\bx_{(i)}^t)_{i=1}^N\}_{t\in\natural}$ %
  be the sequence  generated 
  by \algacronym/, and let  $\avg_{}^t \!\!\triangleq \!\! (1/N)\,\sum_{i=1}^N \bx_{(i)}^t$. 
  Suppose  Assumptions~\ref{ass:cost_functions}, \ref{ass:strong_conn}, \ref{ass:col_stoch}, \ref{ass:T_connectivity}, \ref{ass:surrogate}, and \ref{ass:step-size} are satisfied; 
  then there hold:

  \noindent (i) \texttt{consensus}: $\| \bx_{(i)}^t - \avg_{}^t \| \to 0$ as $t\to \infty$, for all $i\in\until{N}$;

  \noindent (ii) \texttt{convergence}: $\{ \avg_{}^t\}_{t\in\natural}$ is bounded and every of its  limit points  
  is a stationary solution of Problem~\eqref{prob:P}.
  
  \label{thm:convergence}
\end{theorem} 

\algacronym/ enjoys the property that at each iteration 
agents not only solve a low-dimensional optimization problem,
but also transmit a limited amount of information. Moreover, compared to our previous 
scheme in~\cite{notarnicola2017distributed}, in \algacronym/,
the gradient of $f_i$ are computed only with respect to one block rather than the whole 
variable, and this further saves local computation cost.

\section{Numerical Simulations }
\label{sec:simulations}
In this section we test \algacronym/ on an instance of
the sparse regression problem~\eqref{prob:sparse_regression}, where $\KK$
  is a box constraint set, %
  and   $R$ is chosen to be the logarithmic function
  \cite{weston2003use}, with   $\lambda=0.1$; in its DC reformulation in \eqref{eq:DC-reg} we set $\theta=10$ (the specific expression of  $\eta(\theta)$ and thus $r^+$ and $r^-$  therein can be found in \cite{pang2016PartI}). Finally, let $f_i(\mathbf{x})=\|\bb_i - \bD_i \bx\|^2+ R(\mathbf{x})$.%

As  surrogate
$\tilde{f}_{i,\ell}$ of $f_i$ (cf. Assumption \ref{ass:surrogate}), we use 
the linearization  of $f_i$ at the current iterate, i.e.,
\begin{equation}
  \begin{aligned}
  & \tf_{i,\ell}(\bx_{(i,\ell)};\bx_{(i)}^t)
  \\
  = & \left(2\bD_{i,\ell}^\top (\bD_i - \bb_i)\right)^{\!\!\top}\! (\bx_{(i,\ell)} - 
  \bx_{(i,\ell)}^t) \!+\! \frac{\tau_i}{2} \| \bx_{(i,\ell)} - \bx_{(i,\ell)}^t\|^2
  \\
  & - \lambda  \cdot\sum_{k = 1}^{d} \left(  \frac{d r^-  ( ( \bx_{(i,\ell)}^t )_k ) }{dx} \,(\bx_{(i,\ell)} - \bx_{(i,\ell)}^t)_k \right),
  \end{aligned}
\label{eq:sparse_reg_L}
\end{equation}
 where  $(\bx_{(i,\ell)}^t )_k$ denotes the $k$-th scalar component of
 $\bx_{(i,\ell)}^t$, and $dr^{-}(( \bx_{(i,\ell)}^t )_k)/dx$ is the derivative of $r^-$ evaluated at $( \bx_{(i,\ell)}^t )_k$
It is worth noting that~\eqref{eq:sparse_reg_L} admits a unique minimizer, whose expression is omitted because of the space limit. %

We simulated  a network of $N=50$ agents communicating over a fixed undirected 
graph $\GG$, generated using an Erd\H{o}s-R\'enyi random model. 
We compared two  extreme topologies: a densely and a poorly-connected
one, with algebraic connectivity equal to $45$ and $5$, respectively.
There are $500$ optimization variables,  and we set 
$\KK\triangleq [-10,10]^{500}$.    The
components of the ground-truth signal $\bx_0$ are generated independently
according to the Gaussian distribution $\NN(0,1)$.  To impose sparsity on
$\bx_0$, we set the smallest $80\%$ of the entries of $\bx_0$ to zero.  Each
agent $i$ has a measurement matrix $\bD_i \in \real^{50\times 500}$ with
i.i.d. $\NN(0,1)$ distributed entries (with $\ell_2$-normalized rows), and the
observation noise $\bn_i$ has entries i.i.d. distributed according to
$\NN (0,0.5)$.
The diminishing step-size $\gamma^t$ follows the rule $\gamma^t = \gamma^{t-1} (1 - \mu \gamma^{t-1})$,
with $\gamma^0 = 0.1$ and $\mu =10^{-4}$.
The proximal parameter is $\tau_i = 5$ for the poorly connected example
and $\tau_i = 1$ for the densely connected one.

To evaluate the algorithmic performance, we use two merit functions. 
The first one--given by 
$
	J^t \triangleq \norm{ \avg^t -
	\mathcal{P}_{\KK}\big(\SS_{\eta\lambda} \big( \avg^t 
	- (\sum_{i=1}^N \nabla f_i (\avg^t ) -\lambda \cdot \sum_{k=1}^{dB} dr^-((\avg^t)_k)/dx ) \big) \big)}_\infty
$--measures the distance from stationarity
of the average of the agents' iterates $\avg^t$ while the second one--$D^t \triangleq \max_{i\in\until{N}} \| \bx_{(i)}^t - \avg^t \|$--quantifies the consensus disagreement at each iteration.

We compare our algorithm with a non-block-wise distributed gradient
  algorithm; we adapted the gradient-push in
  \cite{nedic2015distributed} to a constrained nonconvex problem according to
  the protocol proposed in \cite{bianchi2013convergence} (no
formal proof of convergence for such scheme  is available in the nonconvex setting).
The performance of \algacronym/ for different choices of the block dimension are
reported in Fig. 1 (a). Recalling that $t$ is the iteration counter used in Algorithm~1, to fairly compare the algorithm runs for different block
sizes, we plot $J^t$ and $D^t$  versus the normalized number of iterations
$t/B$.

The figure shows that for all runs (with different block sizes), both consensus
and stationarity are achieved by \algacronym/ within $100$ normalized
iterations, while the plain gradient scheme using all the blocks is much slower.
Let $t_{\text{end}}$ be the completion time up to a tolerance of $10^{-3}$,
i.e., the iteration counter of the distributed algorithm such that
$J^{t_{\text{end}}} <10^{-3}$. Fig. 1 (b) shows the
normalized completion time $t_{\text{end}} / B$ versus the number of blocks $B$.
This highlights how the communication cost reduces by increasing the number of
blocks.
\begin{figure}[!htbp]
  \centering
      \subfloat[][]{\includegraphics[scale=.94]{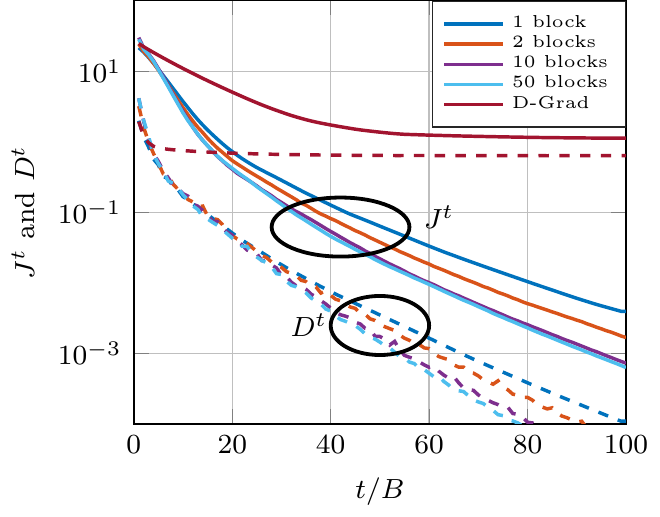}\label{fig:convergence_disagreement_rate}} 
     \subfloat[][]{\includegraphics[scale=.94]{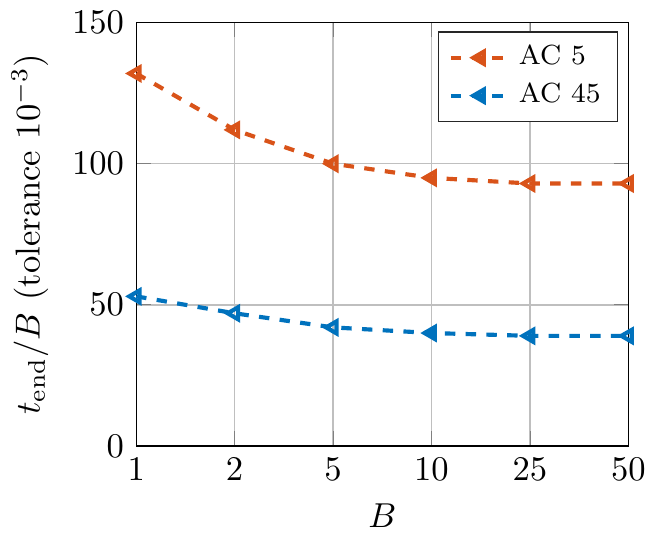}\label{fig:blk_tradeoff}}
\caption{(a) optimality measurement $J^t$ (solid) and consensus error $D^t$ (dashed)
  versus the normalized iteration for several choices of blocks $B$: Algebraic connectivity equal to $5$. (b) Completion time required to 
  obtain $J^t <10^{-3}$ versus the number of blocks $B$.
  }
\end{figure}

\section{Conclusion}
In this paper we studied non-convex distributed big-data
  optimization problems and proposed   \algacronym/  to solve
  them. Leveraging on SCA techniques and  a novel block-tracking/consensus mechanism, the proposed distributed scheme is the fist one unlocking local block-optimization and block-communications. 
Asymptotic convergence to a stationary point of the problem was established, and
numerical tests on the sparse regression problem demonstrated the effectiveness
of algorithm.

\bibliographystyle{IEEEtran}
\bibliography{bigdata_opt_CAMSAP}

\end{document}